\documentclass[lnl,prl,a4paper,twocolumn]{revtex_lnl} 
 \usepackage{epsfig}
 \usepackage{graphicx}
\usepackage[none]{hyphenat} 
\usepackage{pslatex}
\usepackage{xcolor}

\begin{document}


\onecolumngrid
\begin{center}
\begin{large}
{\bf A Statistical Description of Nuclear Reaction Models for Medical Radionuclides: \\
the Paradigmatic Case of $^{47}$Sc Production with Thick Vanadium Targets}
\end{large}
\vspace{.4cm}

M.P. Carante$^1$,  F. Barbaro$^2$, L. Canton$^2$,  A. Colombi$^{1,3}$, A. Fontana$^1$
\vspace{.4cm}

\textit{\small
$^1$ INFN, Sezione di Pavia, Pavia, Italy,
$^2$ INFN, Sezione di Padova, Padova, Italy, \\
$^3$ Dipartimento di Fisica dell'Universit\`a di Pavia, Pavia, Italy.
}
\end{center}
\vspace{0.65cm}

\thispagestyle{empty} 
\pagestyle{empty} 
\twocolumngrid


Theoretical and computational support is an essential ingredient to aid experimental investigations in finding 
efficient routes and conditions for the production of emerging radionuclides for medical applications: this is 
demonstrated for example by the studies  on the scandium, tin and manganese radionuclides described in other contributions 
in this Annual Report \cite{sc43, sc47, sn117, mn52g}. 
Different nuclear reaction codes have been developed and have now become a reference to study the physics of nuclear reactions 
with cyclotrons: however none of them can serve as {optimal tool} and their predictions are often different. 
Therefore, it is important to be able to assess their theoretical variability for a given reaction or for a particular energy interval. 
In this note we focus our attention on the models that are used in code TALYS \cite{talys} (version 1.9): 
this code provides a {default} configuration which has been used in many calculations. On the contrary, some 
authors defined a different configuration of the code parameters (referred often to as \textit{adjusted} \cite{duchemin}) which is believed to provide 
more reliable results in selected cases. In our approach we propose to include all the models provided by the code 
and to consider a  {"Best Theoretical Evaluation" (BTE)} that could be used as a reference value for the models, 
in analogy with the \textit{evaluated} cross section that is recommended for example by IAEA for the experimental data, or with the TENDL library.
\section{Methods}
The nuclear reaction mechanisms relevant for radionuclide production at cyclotrons are dominated by 
compound nucleus formation and by pre-equilibrium emission. 
TALYS is based on the Hauser-Fesbach model where the compound-nucleus formation implies an intermediate 
state of thermodynamic equilibrium before evaporation and de-excitation. 
The formation of this state depends on the level density at the excitation energy corresponding to the projectile incident energy. Regarding pre-equilibrium emission, the models used by TALYS can be divided in two categories, the exciton model and 
the multi-step model, for which the code provides different options.
In general, TALYS has a built-in variety of four models of pre-equilibrium (PE) and six models for nuclear level densities (LD), for a total
of 24 different combinations of models. 

It is important to assess the uncertainties involved in the calculations which depend both on the 
reaction mechanism and on the availability of the relevant nuclear data.
 To achieve this goal, we introduce some simple statistical concepts to deal with
the theoretical variability provided by the different models. Instead of plotting all the 24 curves, 
we compare the different results by means of a statistical band (similarly to what was done in \cite{lamere}).
This band is constructed upon the interquartile range for all models, namely the difference between the third ($Q_3$) and the first ($Q_1$) quartile. 
In this context we define for each energy the BTE cross section 
by taking the average of the first and  third quartile, and associate to it the uncertainty given by the half-width of the  interquartile band:
$$
\sigma_{BTE}=\frac{Q_1+Q_3}{2}\, , \ \ \ \ \ \ \ \ \ \Delta\sigma_{BTE}=\frac{Q_3-Q_1}{2}\, .
$$
This provides a reference value that takes into account the theoretical models variability in a simple way.
The same procedure is applied to evaluate the statistical band not only for cross sections but also for yields, activities, isotopic, and radionuclidic purities, both for the desired nuclide and for its contaminants. If applicable, the band can be plotted also as a function of time to study the time evolution of the relevant quantities.

\begin{figure}[!hbt]
\begin{center}
\includegraphics[width=8cm]{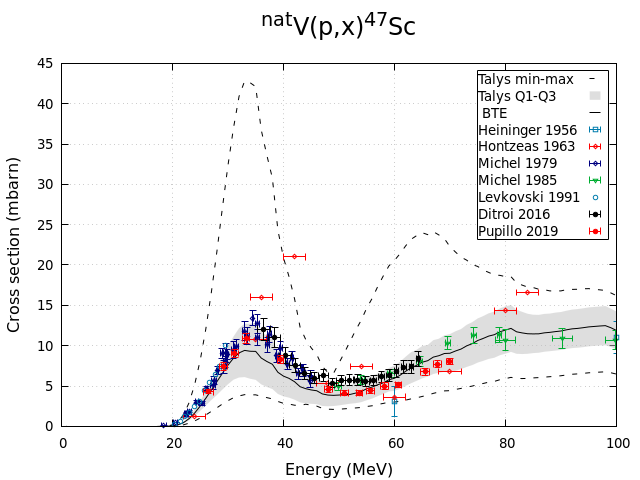}
\end{center}
\caption{Theoretical results for the cross sections with the statistical band description. }
\label{label_fig1}
\end{figure}
\section{Results}
As application of these concepts we consider  
the production of the radionuclide $^{47}$Sc with proton beams impinging on thick targets of  natural Vanadium.
$^{47}$Sc has gained interest recently as an emerging theranostic thanks to the emission of $\beta^-$
particles that can deliver cytotoxic doses to small-medium sized tumors, and $\gamma$-rays suitable for SPECT imaging.
The Vanadium target appears promising due to the possibility to produce $^{47}$Sc with natural materials since their costs are rather inexpensive if compared with alternative, enriched targets. On the other hand, isotopic and radionuclidic purity should be considered of paramount importance in any study on production routes for radiopharmaceuticals application.

We consider, in Fig. 1, the cross section for the $^{nat}V(p,x)^{47}Sc$ reaction. 
The available experimental data, including the ones recently measured within the INFN PASTA project, are compared with this new approach. 
In this case, we considered 3 PE models and 6 different LD models, for a total of 18 combinations. We excluded calculations with the fourth PE model 
because the results turned out unphysical on various occasions. 
The dashed lines indicate the maximum and minimum values of all the models and it is clear that the model variability is too high: depending on the selected models, the description can either over- or under-estimate by large the data. If we consider the interquartile band, spanning quartiles Q1 and Q3, we get a more reasonable description with a narrower band, where only 50\% of the calculations are included. This description appears much more practical, since it keeps the width reasonably small by trimming the calculations at the edge of the set. It can be seen from the figure that the comparison theory-experiment is more meaningful in this way. The solid black line is the BTE and represents the center of the band. The half-thickness is the spread generated by the variability of the models and we suggest it as a theoretical "error" indicator.

Next, we consider the production yield. As well known, this is calculated by an energy integration over the production cross section times the target density and divided by the stopping power. The other factors in front of the integral depend on the physical parameters of the beam and target and they are detailed in the quoted contribution by A. Colombi, et al.
In Fig. 2 we plot the integral yield at the End of Bombardment for an irradiation of 1 h, considering a sufficiently thick target so that the beam energy degradates to zero, i.e. without leaving the target. We calculated the yields for all the 18 TALYS models and repeated the statistical analysis, finding the upper and lower limits of the set of calculations (dashed lines). The narrower interquartile band (in blue) with the BTE given by the center of the band provides a good way to estimate the production yield and includes by construction a simple way to assess the theoretical indeterminations.

The production yield of a target of given thickness can be obtained from the difference of the integral yield (shown in Fig. 2) calculated at $E_{in}$ and $E_{out}$, the energies of the incoming and outgoing beam with respect to the target. If we set $E_{in}=29.5$ and $E_{out}=19.9$ MeV this implies a $^{nat}$V-target thickness of 1080.81 $\mu$m, and the corresponding energy interval is shown in Fig.2 with the green strip. With these irradiation conditions for the $^{47}Sc$ production yield we obtain $Y_p$ $=$  122 $\pm$ 46 KBq/$\mu$Ah with the theoretical uncertainty derived from propagating the width of the band shown in Fig.2. Similarly, we have evaluated the production yield of $^{46}Sc$, the main Scandium contaminant, the only one with an half-life longer than $^{47}$Sc. With the same irradiation condition we get for $^{46}Sc$ $Y_p$ $=$  0.015 $\pm$ 0.006 KBq/$\mu$Ah, and this confirms that the irradiation conditions are optimal for the production with an high level of purity. 
\begin{figure}[!hbt]
\begin{center}
\includegraphics[width=8cm]{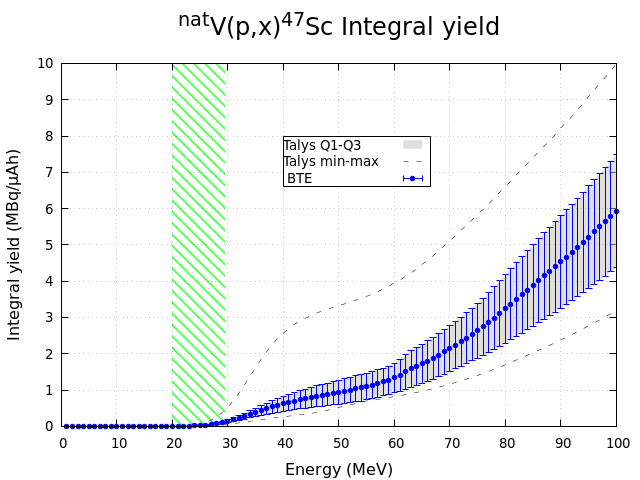}
\end{center}
\caption{Integral yield as a function of beam energy with the error band. }
\label{label_fig1}
\end{figure}

To summarize, we have introduced a tool to describe in a simple and efficient way the outcomes of known nuclear reaction codes which differs from the customary use where tipically a specific single model is selected and the remaining disregarded. The use of the simple statistical procedures, defined herein, allows to overcome this stringent limitation and provides the framework to introduce a more general theoretical evaluation with quantitative uncertainty, constructed on the variability of the built-in theoretical models. Several applications of this technique on the production of emerging radionuclides for medical purposes   
are in progress.


\vspace{-0.5cm}
\begin{thebibliography}{999} 
\bibitem{sc43} L. De Dominicis et al., this LNL Annual Report
\bibitem{sc47} L. Mou et al., this LNL Annual Report
\bibitem{sn117} F. Barbaro et al., this LNL Annual Report
\bibitem{mn52g} A. Colombi et al., this LNL Annual Report
\bibitem{talys} Koning AJ, Hilarie S, Duijvesijn MC, TALYS.1.0 Proceeding of the International Conference on Nuclear Data for Science and Technology, 
April 22-27, 2007, Nice, France, EDP Sciences (2008) 211
\bibitem{duchemin} Duchemin C. et al., Phys Med Biol DOI 10.1088/0031-9155 /60/3/931 (2015)
\bibitem{lamere} E. Lamere et al., Phys. Rev. C 100, 034614 (2019)
\end{thebibliography}
\end{document}